\documentclass[preprint2]{aastex}

\shorttitle{Bica \& Dutra } \shortauthors{Census of SMC clusters}

\newcommand\acta{{Acta~Astron.}}%

\begin{document}


\title{Updating the census of  star clusters
\\ in the Small Magellanic Cloud.}

\author{E. Bica }
\affil{Universidade Federal do Rio Grande do Sul, Instituto de
Fisica, CP 15051, Porto Alegre 91501-970, RS, Brazil}
\email{bica@if.ufrgs.br}

 \and

\author{C. M. Dutra }
\affil{Universidade Federal do Rio Grande do Sul, Instituto de
Fisica, CP 15051, Porto Alegre 91501-970, RS, Brazil}
\email{dutra@if.ufrgs.br}

\begin{abstract}
Surveys using CCD detectors are retrieving bright and faint cataloged 
clusters and revealing new ones in the Magellanic Clouds. 
This paper discusses  the  contribution of the  OGLE 
Survey to the overall census of star clusters in the SMC.
A detailed cross-identification indicates that the new objects in the 
SMC OGLE catalog are 46. The increase in the number of cataloged clusters
is $\approx 7\%$, the total sample being $ \approx 700$. This updated census includes embedded clusters in HII regions and a density range attaining loose systems.

\end{abstract}


\section{Introduction}

The Magellanic Clouds contain rich star cluster systems
\citep{hod86,hod88}. The distances of the Clouds and their rather
high galactic latitudes make them ideal targets to probe total
populations of extended objects.

 It is becoming possible to map
out the overall angular distributions of star clusters,
associations and HII regions, see e.g. the revisions of previously
cataloged and newly identified objects in the SMC and LMC
\citep{bic95,bic99}. Such distributions help one better
understand star formation mechanisms and the evolution of the
Magellanic System. The homogeneous surveys above were carried out on
ESO/SERC  R and J Sky Survey Schmidt Plates.

 CCD survey results are becoming available which provide deep images in particular areas. A sector of the LMC was studied by \citet{zar97} using UBVI filters and they identified previous and new clusters. These objects were cross-identified in detail with the previous literature in \citet{bic99}, and were included in that catalog.  Recently, \citet{pie98} built a catalog of SMC clusters from the Optical Gravitational Lensing Experiment (OGLE) BVI database. The region covered by the OGLE Survey is $\approx$ 2.4 square degrees in the central parts of the SMC. They reported 238 clusters and presented a cross-identification concluding that 72 clusters were newly cataloged.

In the present paper we perform a detailed
cross-identification of the objects in the OGLE catalog with
those in previous works,  homogenizing classifications. We indicate intrinsically new objects which in
turn have implications on the star cluster census. According to
\citet{hod86} the total cluster population in the SMC would be
$\approx$ 900 if it were surveyed entirely with 4m telescope deep B plates as he did for a selection of fields. Considering
incompleteness effects related to faint turnoffs he estimated a grand total of $\approx$
2000.  Since Hodge's (1986) study new objects have been identified in the SMC \citep{bic95,pie98} and it is important to update the census in view of shedding light on the SMC history of star cluster formation and dissolution. In Section 2 we discuss the cross-identification procedures
and present the results. In Section 3 we compare the angular
distribution of the SMC\_OGLE objects with that of all extended
objects in the SMC, and discuss the impact of the new SMC\_OGLE
objects on the SMC census of extended objects, in particular star
clusters. In Section 4 we give the concluding remarks.

\section{Cross-identifications}

\citet{pie98} presented the catalog of OGLE objects in the SMC, together with
 I-Band CCD images of
each object. We used the information therein provided, particularly
the coordinates and images for the present cross-identification
with the objects in Bica \& Schmitt (1995, hereafter BS95).

We overplotted the 238 SMC\_OGLE objects on J2000 maps containing
the objects in BS95. The objects resulting very close in position
had their CCD images in \citet{pie98} compared to the corresponding
fields in the ESO/SERC  R and J Schmidt plates and Digitized Sky
Survey (DSS) images, in order to check equivalences, which as rule
occurred. Isolated objects turned out to be new objects.

We measured diameters and position angles of the
new SMC\_OGLE objects and classified them homogeneously with BS95 and
\citet{bic99}. The object types in the latter classifications are:
C for star cluster, A for emissionless association, CA and AC for
objects with intermediate properties, NA for HII regions and
embedded associations, NC for HII regions and embedded star
clusters or high surface brightness compact HII regions, N for
supernova remnants, AN and CN are respectively associations and
clusters which show traces of emission.

The results for all SMC\_OGLE objects following the BS95 and
\citet{bic99} catalog  format are given in Table 1. By columns:
(1) The Sky Survey field quadrant where the object is best seen.
(2) Object cross-identification in the different catalogs.  (3)
and (4) are right ascension and declination for the epoch 2000
respectively, (5) Object Type.(6) and (7) Major and minor
diameters respectively. (8) Position angle of major axis
(0$^{\circ}$=N, 90$^{\circ}$=E). (9) Remarks: `mP', `mT' indicate
member of pair, triple  etc, respectively; `\&' indicates
additional designations to column 2.

We confirmed most of the  cross-identifications by \citet{pie98}
for the 166 objects therein indicated as having previous identifications. However  of the 72
SMC\_OGLE objects reported as newly found  by \citet{pie98} we concluded that  26 had previous identification in the literature, while 46
are intrinsically new objects. These 46 objects are relatively
isolated, and their typical appearance on the DSS images is a few
enhanced pixels, basically unresolved, thus objects  which indeed
required deep CCD images to be recognized as clusters.

 Two OGLE objects are duplicated,
SMC\_OGLE236 = SMC\_OGLE144 and
 SMC\_OGLE175  = SMC\_OGLE19 (Table 1). Two single objects in the OGLE catalog, SMC\_OGLE26 and SMC\_OGLE33, were considered to be pairs in BS95 (the north and south components of NGC248 and H86-78). Adopting the latter separations the total number of objects with SMC\_OGLE designation remains 238 as in \citet{pie98}. One case of close position between an SMC\_OGLE object (SMC\_OGLE31) and one in BS95 (B36) resulted in two different objects. The coordinates of SMC\_OGLE166 in \citet{pie98} did not correspond to the cluster image; the correct coordinates were kindly provided by Dr. Pietrzy\'nski.

Table 2 shows the distribution  of SMC\_OGLE objects in different
catalogs considering the first chronological designation. Column
1 shows the acronym; column 2 the catalog reference; column 3
counts made in Pietrzy\'nski et al.'s (1998) Table 2; finally in
column 4 counts made in Table 1 of the present work.  This
comparison basically shows the distribution of the difference
between the 72 initially reported as new and the 46 intrinsically
new SMC\_OGLE objects. This difference arises mainly from faint
clusters in the B \citep{bru76}, H86 \citep{hod86} and BS (BS95) catalogs.

Concerning the distributions in Table 2, note  that the L61
\citep{lin61} and  MA \citep{mey93} catalogs deal with emission
stellar sources, but containing some extended objects which were
included in BS95. These extended emission line objects are HII
Regions and embedded star clusters. The L61 designations given to
seven SMC\_OGLE objects in Table 2 of \citet{pie98} have SMC-N
\citep{hen56} as first chronological identification in BS95. Note
also that one SMC\_OGLE object has a counterpart in the MA
catalog.

\subsection{Updated Electronic Version of the BS95 Catalog}

We provide in Table 3  the first 5 lines of an updated electronic
version of the revised and extended catalog of star clusters,
associations and emission nebulae in the SMC/Bridge (BS95). This
incorporates the present results concerning the SMC\_OGLE objects
(Table 1), in particular the 46 new entries. We note that the
equatorial coordinates in the present version are J2000, while in
BS95 they were B1950. As pointed out by BS95 an updated catalog condensing the literature results is very useful for future surveys.

 Outside the OGLE survey area we include 3 objects which were previously in the list of excluded catalog entries of BS95 (their Table 3) because they were not clearly interpreted as clusters in the ESO/SERC Schmidt Plates. These objects are: (i) B133 \citep{bru76}, which appears to be a physical system \citep{kon80,hod83}; (ii) H86-95 and H86-96 indicated as a cluster pair in \citep{hat90}, which are probably clusters as seen in the second generation Digitized Sky Survey images.

The SMC/Bridge catalog now totals 1237 objects: 595 classified
as clusters (C+CA+CN), 350 as associations (A+AC+AN) and 292
related to emission nebulae (N+NC+NA). Considering also the recent LMC
catalog \citep{bic99} the number of extended objects in the
Magellanic System is now 7895.

\section{Angular distribution and census}

In the following discussions we adopt the updated BS95 catalog
(Table 3), which incorporates that of the OGLE survey (Table 1). We
consider as SMC objects those with right ascension less than
$2^h$, thus excluding the Bridge region.

Figure 1 shows the angular distribution of all SMC\_OGLE objects
compared to that of all extended objects in the updated BS95
catalog. The OGLE survey covers the central regions of the SMC,
including the very dense bar region to the southwest and the dense
extension to the northeast. Note that there are moderately dense
fields around the OGLE survey area in which future CCD surveys
will certainly detect new objects. Also  the low density outer
zones and the SMC Wing region to the southeast are important to be
surveyed.

Figure 2 superimposes the angular distribution of the 46
intrinsically  new  SMC\_OGLE objects to that of all (238) objects
in the SMC\_OGLE catalog. The new objects are more frequent in
the Bar region, but they are also numerous in the northeast
extension.

In Table 4 we show distributions of object types in the SMC for
different spatial extractions in the updated BS95 and OGLE
catalogs. In column 2 (whole SMC) there are 1122 objects
distributed among (clusters : associations : nebulae) as (584 :
252 : 286). The 46 new  OGLE objects and the three catalog
additions outside the OGLE region (Section 2.1) increased the
number of SMC objects by $\approx$ 5\%.

We also show in Table 4 object type counts occurring in the OGLE
survey area: in  column 3 all objects from the updated BS95
catalog, in column 4 all objects in the revised OGLE catalog
(Table 1), and finally in column 5 the intrinsically new OGLE
objects. We compute 631 extended objects in the OGLE survey area
as compared to  238 SMC\_OGLE objects. The large OB associations
and nebular complexes explain this difference in part (mostly
included in the A and NA types), but there occur many clusters (C
type) in the OGLE survey area not included in the OGLE catalog.
The classification C, CA to AC is one of decreasing density
\citep{bic99} and most of the new OGLE objects (column 5) were
classified in the CA type. Finally we point out that two new
OGLE objects are related to emission (NC and NA types), and
$\approx$ 10 \% of the objects in the OGLE catalog (column 4)
are related to emission or have traces of emission (CN type). The
emission is better seen in ESO/SERC R plates owing to H$\alpha$
than in the I-band CCD images.

We show in column 6 of Table 4 the counts for all objects outside
the OGLE survey area, and in column 7 a crude prediction of
objects that may be detected by similar CCD surveys. For each
object type we computed the fraction of new OGLE objects (column
5) with respect to all objects of the same type in the area
(column 3). It is expected that the average age in the outer parts
be larger than in the inner parts. This effect would increase the
number of new CCD objects in the outer parts since the turnoffs
would be fainter, and also because crowding effects are less
important. On the other hand the latter point does not favor new
detections because many of these relatively old objects would have
already been detected in previous photographic surveys. It is also
worth noting that the older age effect towards outer regions is
not isotropic, since young clusters certainly occur to the
southeast (Wing and Bridge regions), which favors more previous
detections. Under the assumption that such effects tend to
compensate, the number of new objects in the outer region would be
$\approx$ 42 (column 7).

 A cluster census depends on the definition of cluster itself. One can include embedded clusters in 
 HII regions and extend the density range to loose systems, as done for the LMC \citep{bic99}.  
 Considering the census of SMC star clusters by including a density range (C, CA and AC types) the actual number is 633 and including predictions 672. Considering also the clusters related to emission (NC and CN types) the actual cataloged number is 719 and including predictions 759. With respect to a cluster population of 719 the new objects in the OGLE catalog imply an increase of $\approx 7\%$.

 \citet{hod86} carried out a similar analysis by comparing the number of clusters cataloged in deep 4m telescope B plates in selected areas to that of known clusters in SMC catalogs at that time. He predicted $\approx$ 900 clusters if all the SMC were surveyed with similarly deep plates, and $\approx$ 2000 clusters if small older clusters were detectable. The plate limits in \citet{hod86} are $B = 23$ and $B = 22$, respectively in the outer and core SMC regions, while in the OGLE survey they are slightly less deep with $B \approx 21.2$, $V \approx 21.5$ and $I \approx 21.0$ \citep{uda98}. The present number of cataloged clusters in Table 3 (719) is still short of $\approx$ 180 with respect to  Hodge's prediction for the detection level of the 4m plates (900), but it is considerably larger than that actually known at that time ($\approx 600$).

Hodge's (1986) 4m telescope photographic survey is to date still the deepest one in selected areas in the SMC. Note that 53 faint cluster candidates (indicated by Hodge as probable or questionable clusters) 
remain in Table 3 of BS95 owing to limitations in the ESO/SERC Schmidt plates. 
This number excludes H86-95 and H86-96 (now in the present catalog in Table 3 
- see Section 2.1) and the duplications H86-131=128, 137=133, 161=158 and 
168=165 (now also in Table 3). Such faint clusters may turn out to be crucial
 to infer the rate whereby intermediate age and old clusters formed in possible
  bursts and/or dissolved in the SMC and also LMC \citep{gei97}. 
  The brightest star in these faint clusters is  mostly in the range
   $18 \le$ B $\le 22$ \citep{hod86} which places them as candidate 
   intermediate/old age clusters since some bright red giant/clump stars 
   may be present or not in such underpopulated objects. 
   The turnoff magnitude for old clusters in the SMC
    is B $\approx 22-23$ \citep{hod86}. Assuming the faint objects as clusters the updated sample
     would be 772 (719 + 53), still short of $\approx 130$ with respect to the 4m B plate survey predictions for the whole SMC with comparable plate limits.     

We conclude that deep CCD surveys like  OGLE  will certainly
reveal new clusters in the SMC intermediate and outer parts, and
deeper surveys would be necessary (especially in the central
regions) to attain $\approx$ 900 objects. As an example of
detection of very faint clusters by means of deep localized
images, see the recent discovery of two clusters in crowded LMC
fields with HST \citep{san98}. Finally, deep field CCD surveys
would be necessary to check the existence of small old (and
intermediate age) clusters in order to attain a total population
of $\approx$ 2000. One possibility is that most or part of such
older low mass objects have dissolved.

\section{Concluding remarks}

 The updating of the SMC extended object catalog by including the OGLE survey and some 
 additional results was carried out.
The 46 new OGLE objects increase  the number of extended objects in the SMC by
  $\approx$ 5\% and star clusters themselves by $\approx$ 7\% . If a similar CCD survey 
  were carried out in the whole SMC area, a simple estimate suggests that
   $\approx$ 40 additional objects could be detected. The present number of
    star clusters (considering also those related to emission and loose systems) is 719, still
     short by $\approx$ 180 with respect to Hodge's (1986) prediction for a global survey in the SMC attaining $B = 23$ and $B = 22$ in the outer parts and core respectively. Deeper field CCD surveys would be necessary to check the existence of small old (and intermediate age) clusters in order to attain a grand total population of $\approx$ 2000. However it is possible that most or part of such older low mass objects have dissolved.

\acknowledgments

 We acknowledge the Brazilian institution CNPq for
support.


\clearpage



\figcaption[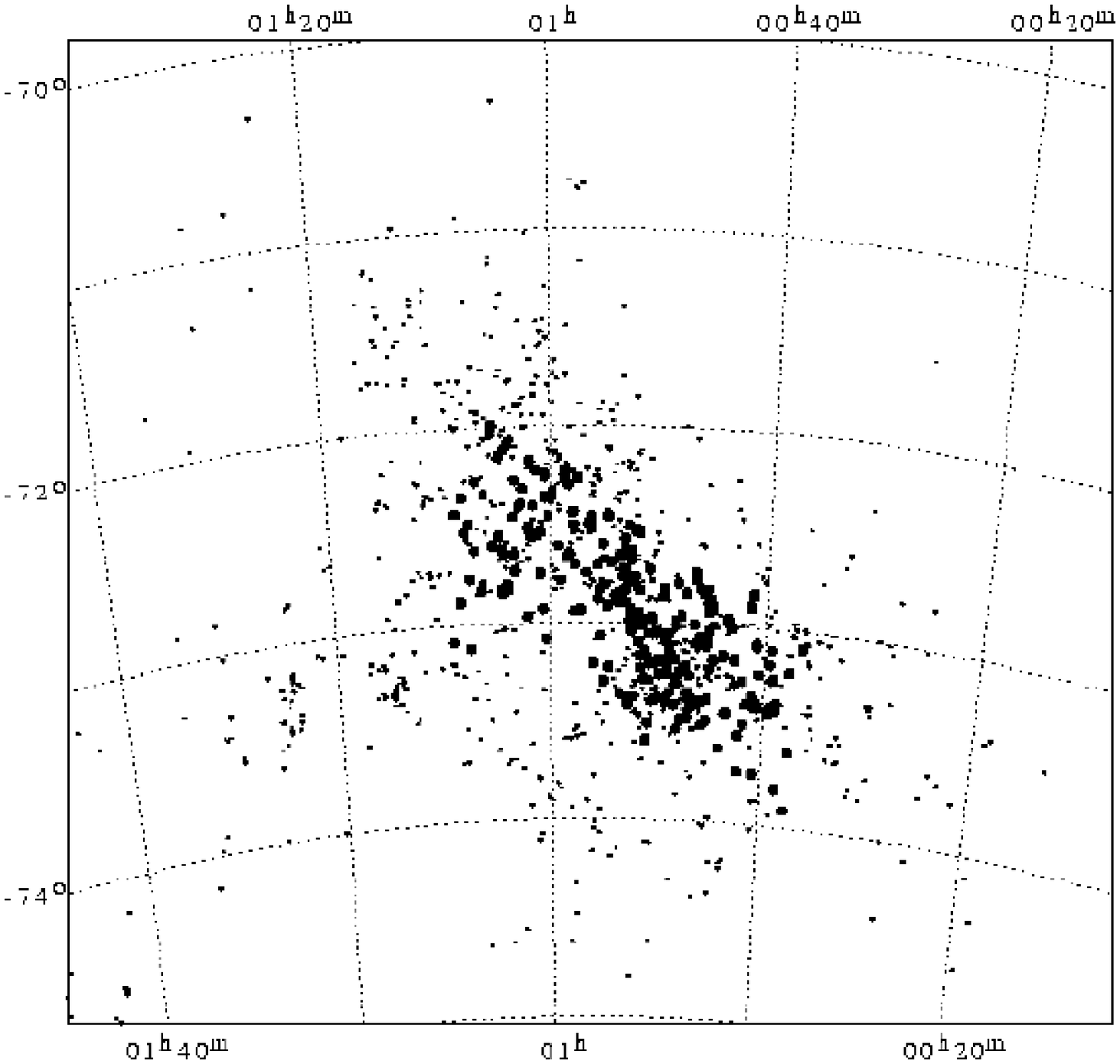]{Angular distribution (J2000 $\alpha$,
$\delta$) of the 238 SMC\_OGLE objects (large dots) superimposed
on that of the SMC objects (small dots) from BS95. \label{fig1}}

\figcaption[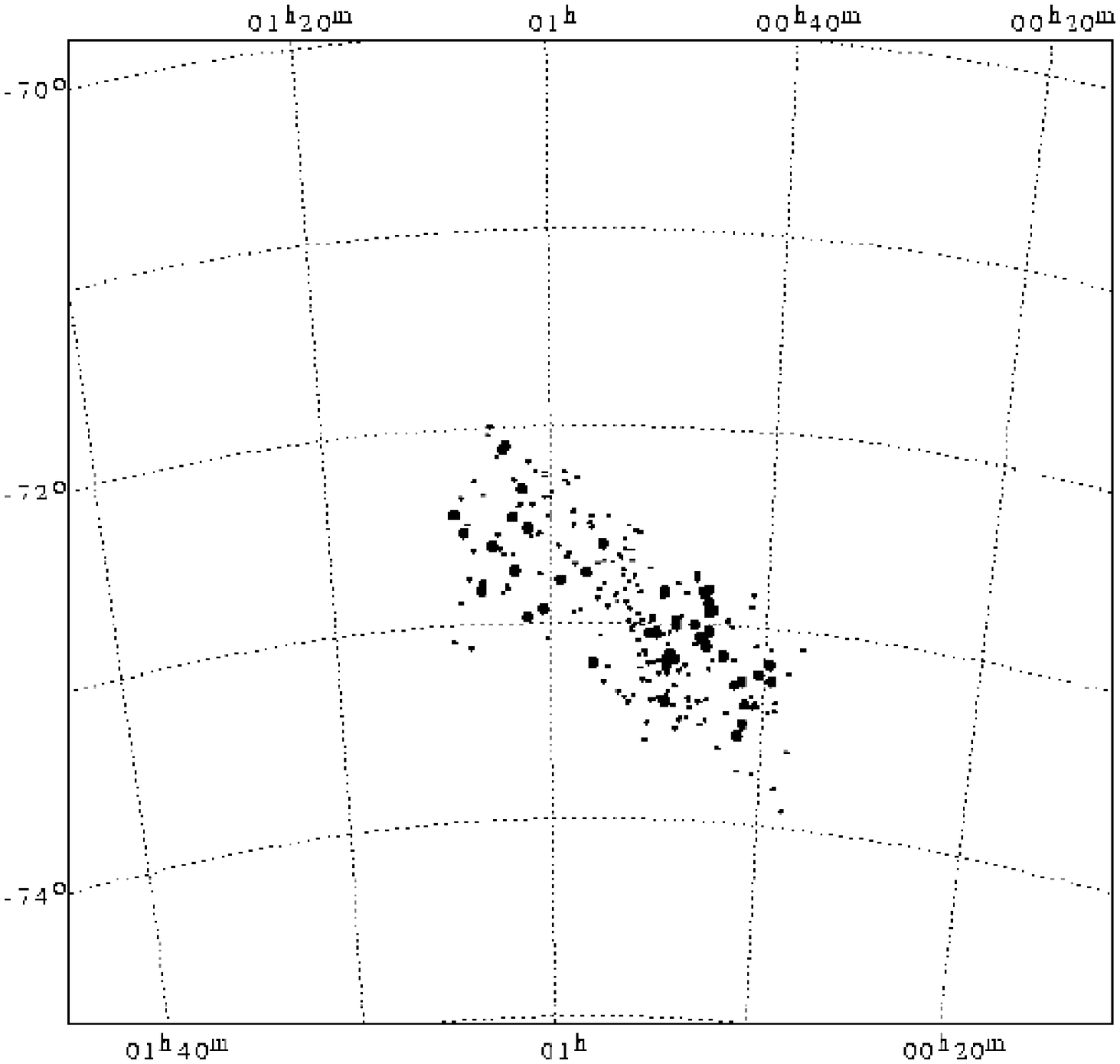]{Angular distribution (J2000 $\alpha$,
$\delta$) of the 46 intrinsically new SMC\_OGLE objects (large
dots) superimposed on that of the 238 SMC\_OGLE objects (small
dots). \label{fig2}}





\clearpage

\begin{table*}
\caption[]{Presently cross-identified SMC-OGLE catalog}
\begin{scriptsize}
\renewcommand{\tabcolsep}{0.9mm}
\begin{tabular}{llccccccl}
\tableline\tableline
 29nw&H86-35,OGLE1              &0:36:38&-73:05:09& C&0.35&0.35& - &                                                   \\
 29nw&HW11,OGLE2            &0:37:33&-73:36:43& C&1.30&1.30& - &                                                   \\
 29nw&L19,OGLE3             &0:37:42&-73:54:30& C&1.70&1.70& - &                                                   \\
 29nw&B10,OGLE4             &0:37:44&-73:12:40& C&0.80&0.80& - &                                                   \\
 29nw&B14,OGLE162           &0:38:37&-73:48:21& C&0.60&0.45&160&                                                   \\
 29nw&HW12,OGLE163          &0:38:51&-73:22:27& C&0.80&0.80& - &                                                   \\
 29nw&H86-48,OGLE164        &0:38:56&-73:24:32& C&0.50&0.40& 30&                                                   \\
 29nw&BS14,OGLE165          &0:39:12&-73:14:46& C&0.55&0.55& - &mP                                                 \\
 29nw&OGLE5                 &0:39:22&-73:15:28&CA&0.85&0.75& 90&mP                                                 \\
 29nw&H86-55,OGLE167        &0:39:26&-73:06:23& C&0.45&0.35& 80&                                                   \\
 29nw&HW13,OGLE168          &0:39:31&-73:25:26& C&0.75&0.60& 80&                                                   \\
 29nw&OGLE6                 &0:39:33&-73:10:37& C&0.80&0.80& - &                                                   \\
 29nw&OGLE7                 &0:40:30&-73:13:50&CA&0.95&0.95& - &                                                   \\
 29nw&NGC220,K18,L22,ESO29SC3,   &0:40:31&-73:24:10& C&1.20&1.20& - &mT,in H-A3  \& OGLE8                           \\
 29nw&B26,OGLE169           &0:40:45&-73:44:26& C&0.90&0.90& - &                                                   \\
 29nw&NGC222,K19,L24,ESO29SC4,   &0:40:44&-73:23:00& C&1.20&1.20& - &mT,in H-A3  \& OGLE9                           \\
 29nw&H86-62,OGLE10         &0:40:48&-73:05:17& C&0.60&0.60& - &mT                                                 \\
 29nw&B23,OGLE170           &0:40:55&-73:24:07& C&0.60&0.50& 40&mT,in H-A3                                         \\
29nw &NGC231,K20,L25,ESO29SC5,    &0:41:06&-73:21:07& C&1.80&1.80&
- &mP,in H-A3  \& OGLE11                          \\
 29nw&B21,OGLE171           &0:41:15&-72:49:55& C&0.35&0.35& - &mP                                                 \\
 29nw&K21,L27,OGLE12        &0:41:24&-72:53:27& C&2.50&2.50& - &                                                   \\
 29nw&OGLE172               &0:41:48&-73:23:27& C&0.20&0.20& - &                                                   \\
 29nw&OGLE166               &0:41:56&-73:29:16& C&0.60&0.50& 80&                                                   \\
 29nw&OGLE173               &0:42:12&-73:16:02& C&0.20&0.20& - &                                                   \\
 29nw&HW16,OGLE13           &0:42:22&-73:44:03&CN&0.60&0.60& - &mP,in SMC-DEM7                                     \\
 29nw&OGLE14                &0:42:28&-73:32:50&AC&0.80&0.60&170&                                                   \\
 29nw&OGLE15                &0:42:54&-73:17:37&CA&1.00&1.00& - &in? sup? H-A4                                      \\
 29nw&BS16,OGLE16           &0:42:58&-73:10:07& C&0.55&0.50& 10&in H-A5                                            \\
 29nw&BS17,OGLE174          &0:43:14&-73:00:43&CA&0.80&0.65& 60&                                                   \\
 29nw&NGC241,K22w,L29w,         &0:43:33&-73:26:25& C&0.95&0.95& - &mP \& ESO29SC6w,OGLE17                          \\
 29nw&NGC242,K22e,L29e,         &0:43:38&-73:26:37& C&0.75&0.75& - &mP \& ESO29SC6e,BH1,OGLE18                      \\
 29nw&B31,OGLE19,OGLE175&0:43:38&-72:57:31& C&0.50&0.40&150&mT                                                 \\
 29nw&BS20,OGLE20           &0:43:38&-72:58:48& C&0.45&0.45& - &mT                                                 \\
 29nw&H86-70,OGLE21         &0:43:44&-72:58:36& C&0.65&0.45& 50&mT                                                 \\
 29nw&OGLE22                &0:43:58&-73:09:08&CA&0.75&0.75& - &                                                   \\
 29nw&B33,OGLE23            &0:44:13&-73:37:08& C&0.50&0.50& - &                                                   \\
 29nw&B34,OGLE176           &0:44:52&-73:00:07& C&0.60&0.60& - &br* edge                                           \\
 29nw&BS27,OGLE177          &0:44:55&-73:10:27& C&0.40&0.35& 80&mP,in H86-72                                       \\
 29nw&OGLE24                &0:45:01&-72:55:17& A&1.90&1.90& - &                                                   \\
 29nw&BS28,OGLE178          &0:45:11&-72:52:31&CA&0.70&0.50& 60&                                                   \\
 29nw&H86-74,OGLE25         &0:45:14&-73:13:09& C&0.70&0.60& 50&in SMC-DEM13                                       \\
 29nw&OGLE179               &0:45:21&-73:02:08&CA&0.50&0.50& - &                                                   \\
 29nw&NGC248n,SMC-N13B,L61-67n,  &0:45:24&-73:22:34&NA&0.70&0.60&110&mP \& SMC-DEM16n,ESO29EN8n,MA101,OGLE26n        \\
 29nw&OGLE180               &0:45:23&-72:55:43& C&0.25&0.25& - &                                                   \\
 29nw&NGC248s,SMC-N13A,L61-67s,  &0:45:26&-73:23:04&NC&0.70&0.55&150&mP \& SMC-DEM16s,ESO29EN8s,MA103,OGLE26s        \\
 29nw&B39,OGLE27            &0:45:26&-73:28:53& C&0.55&0.55& - &mP                                                 \\
 29nw&OGLE28                &0:45:28&-72:49:10& C&0.40&0.40& - &                                                   \\
 29nw&OGLE181               &0:45:28&-72:53:09&CA&0.70&0.50& 60&                                                   \\
 29nw&NGC249,B35,ESO29EN9,       &0:45:30&-73:04:43&NA&1.10&0.70&  0&in SMC-N12B \& OGLE29                           \\
 29nw&OGLE30                &0:45:33&-73:06:27&CA&0.75&0.75& - &                                                   \\
 29nw&OGLE31                &0:45:51&-72:50:25&CA&0.55&0.45& 90&mP, not B36                                        \\
 29nw&NGC256,K23,L30,ESO29SC11,  &0:45:54&-73:30:24& C&0.90&0.90& - &\& OGLE32                                       \\
 29nw&H86-76,OGLE182        &0:46:02&-73:23:44& C&0.45&0.45& - &mT,in SMC-DEM21                                    \\
 29nw&H86-80,OGLE184        &0:46:11&-72:49:03& C&0.35&0.35& - &                                                   \\
 29nw&H86-78n,OGLE33n       &0:46:12&-73:23:27&CN&0.45&0.45& - &mT,in SMC-N16                                      \\
 29nw&OGLE183               &0:46:10&-73:03:56&CA&0.35&0.30&160&                                                   \\
 29nw&H86-78s,OGLE33s       &0:46:12&-73:23:39&CN&0.45&0.40& 60&mT,in SMC-N16                                      \\
 29nw&NGC261,B42,ESO29EN12,      &0:46:32&-73:05:55&NA&1.30&1.10&100&in SMC-N12A  \& OGLE34                          \\
 29nw&L31,OGLE36            &0:46:35&-72:44:32& C&1.10&0.85& 30&mT                                                 \\
 29nw&H86-83,OGLE35         &0:46:34&-72:46:26& C&0.70&0.70& - &mT                                                 \\
 29nw&H86-84,OGLE185        &0:46:34&-72:45:56& C&0.40&0.40& - &mT                                                 \\
 29nw&OGLE37                &0:46:41&-73:00:00& C&0.80&0.70& 10&                                                   \\
 29nw&H86-85,OGLE186        &0:46:56&-73:25:25& C&0.60&0.50&150&in SMC-DEM29                                       \\
 29nw&H86-86,OGLE40         &0:47:01&-73:23:35& C&0.80&0.65&110&mP,in H-A9                                         \\
\tableline
\end{tabular}
\end{scriptsize}
\tablecomments{Units of right ascension hours, minutes and seconds, and declination degrees, arcminutes and arcseconds.
}
\end{table*}

\clearpage

\begin{table*}
\begin{scriptsize}
\renewcommand{\tabcolsep}{0.9mm}
\begin{tabular}{llccccccl}
\tableline\tableline &&&&&D$_{max}$&D$_{min}$&P.A.&\\
Plate&Name&R.A.(2000)&Dec(2000)&T&(arcmin)&(arcmin)&(deg)&Remarks\\
(1)&(2)&(3)&(4)&(5)&(6)&(7)&(8)&(9)\\ \tableline
 29nw&H86-89,OGLE38         &0:47:06&-73:15:24& C&0.85&0.65&150&                                                   \\
 29nw&H86-87,OGLE187        &0:47:06&-73:22:17& C&0.80&0.70& 90&mP,in H-A9                                         \\
 29nw&NGC265,K24,L34,ESO29SC14,  &0:47:12&-73:28:38& C&1.20&1.20& - &\& OGLE39                                       \\
 29nw&L33,OGLE41            &0:47:25&-72:50:27& C&1.00&0.80&130&                                                   \\
 29nw&H86-93,MA172,OGLE188  &0:47:24&-73:12:20&CN&0.40&0.40& - &in H-A11                                           \\
 29nw&BS35,OGLE42           &0:47:50&-73:28:42& C&0.70&0.70& - &mP                                                 \\
 29nw&H86-97,OGLE43         &0:47:52&-73:13:20& C&0.70&0.60& 60&                                                   \\
 29nw&H86-98,OGLE44         &0:47:55&-72:57:20&CA&0.90&0.65& 60&                                                   \\
 29nw&K25,L35,OGLE45        &0:48:01&-73:29:10& C&1.20&1.20& - &mP                                                 \\
 29nw&SMC-N25,L61-106,SMC-DEM38,&0:48:09&-73:14:19&NA&0.85&0.85& - &mT \& MA208,OGLE189                             \\
29nw&H86-99,OGLE190        &0:48:13&-72:47:35&CA&0.65&0.65& -
&mP                                                 \\
29nw&H86-99,OGLE190        &0:48:13&-72:47:35&CA&0.65&0.65& -
&mP                                                 \\
 29nw&H86-100,OGLE191       &0:48:20&-72:47:42&CA&0.75&0.75& - &mP                                                 \\
 29nw&NGC269,K26,L37,ESO29SC16,  &0:48:21&-73:31:49& C&1.20&1.20& - &\& OGLE46                                       \\
 29nw&OGLE192               &0:48:26&-73:00:26& C&0.35&0.35& - &                                                   \\
 29nw&OGLE47                &0:48:28&-72:59:00&AC&1.20&1.20& - &                                                   \\
 29nw&B47,OGLE48            &0:48:33&-73:18:25& C&1.00&1.00& - &                                                   \\
 29nw&B48,OGLE49            &0:48:37&-73:24:53&CA&1.30&1.10&140&in H-A16                                           \\
 29nw&OGLE193               &0:48:37&-73:10:45&CA&0.50&0.50& - &                                                   \\
 29nw&OGLE50                &0:48:59&-73:09:04&NC&0.75&0.75& - &in SMC-N30                                         \\
 29nw&H86-103,OGLE51        &0:49:05&-73:03:04& C&0.55&0.45& 40&                                                   \\
 29nw&BS41,OGLE194          &0:49:06&-73:21:10& C&0.55&0.55& - &mT                                                 \\
 29nw&H86-104,OGLE52        &0:49:12&-73:06:31& C&0.40&0.40& - &                                                   \\
 29nw&BS42,OGLE195          &0:49:16&-73:14:57&CA&1.00&1.00& - &in SMC-DEM49                                       \\
 29nw&L39,OGLE54            &0:49:18&-73:22:20& C&0.70&0.55&170&mT,in BS43                                         \\
 29nw&OGLE53                &0:49:18&-73:12:42& C&1.00&0.80& 10&                                                   \\
 29nw&OGLE55                &0:49:21&-73:11:02& C&0.60&0.60& - &                                                   \\
 29nw&OGLE196               &0:49:27&-73:23:55& C&0.35&0.35& - &                                                   \\
 29nw&OGLE56                &0:49:36&-72:50:13&CA&0.80&0.60&100&mT,in SMC-DEM46e                                   \\
 29nw&B52,OGLE57            &0:49:40&-73:03:32& C&1.10&1.10& - &in SMC-DEM51                                       \\
 29nw&H86-109,OGLE58        &0:49:45&-72:51:58& C&0.45&0.45& - &mT                                                 \\
 29nw&H86-107,OGLE61        &0:50:00&-73:15:18&CA&1.20&0.75&130&in SMC-DEM49                                       \\
 29nw&B53,OGLE197           &0:50:04&-73:23:04& C&0.95&0.95& - &mP                                                 \\
 29nw&H86-112,OGLE198       &0:50:08&-73:11:26& C&0.65&0.55&  0&in H-A18                                           \\
 29nw&OGLE199               &0:50:15&-73:03:15&CA&0.25&0.25& - &mP, in? sup? SMC-DEM51                             \\
 29nw&BS45,OGLE59           &0:50:16&-73:02:00&CA&1.00&0.90& 70&mP,in SMC-DEM51                                    \\
 29nw&B55,OGLE60            &0:50:22&-73:23:16& C&0.70&0.60&110&mP                                                 \\
 29nw&B54,OGLE62            &0:50:28&-73:12:12& C&0.65&0.65& - &                                                   \\
 29nw&H86-115,OGLE63        &0:50:37&-73:03:28&AC&1.60&1.20& 40&mP,in SMC-DEM51                                    \\
 29nw&BS46,OGLE200          &0:50:39&-72:58:44& C&0.50&0.45& 60&mP,in L40                                          \\
 29nw&H86-116,OGLE64        &0:50:40&-72:57:55& C&0.50&0.50& - &mP,in L40                                          \\
 29nw&BS48,OGLE201          &0:50:42&-73:23:49&AC&0.85&0.55& 80&mP,in SMC-DEM53                                    \\
 29nw&L41,OGLE67            &0:50:56&-72:43:40& C&0.65&0.65& - &in H-A25                                           \\
 29nw&OGLE65                &0:50:55&-73:03:27& C&0.65&0.65&-  &mP                                                 \\
 29nw&B56,OGLE66            &0:50:55&-73:12:11& C&0.45&0.45& - &                                                   \\
 29nw&BS40,OGLE68           &0:50:56&-73:17:21&CA&0.90&0.80&140&in H-A24                                           \\
 29nw&H86-105,OGLE202       &0:50:59&-73:30:13& C&0.45&0.45& - &                                                   \\
 29nw&NGC290,L42,ESO29SC19,      &0:51:14&-73:09:41& C&1.10&1.10& - &in BS51  \& OGLE69                              \\
 29nw&H86-113,OGLE203       &0:51:10&-73:35:24& C&0.40&0.40& - &                                                   \\
 29nw&H86-121,OGLE204       &0:51:21&-73:08:19& C&0.55&0.55& - &in BS51                                            \\
 29nw&BS251,OGLE70          &0:51:26&-73:17:00&CA&0.40&0.35&140&                                                   \\
 29nw&H86-124,OGLE205       &0:51:32&-72:58:45& C&0.85&0.65& 80&                                                   \\
 29nw&B57,OGLE71            &0:51:32&-73:00:38& C&1.20&1.20& - &                                                   \\
 29nw&SMC-N45,L61-189,B60,      &0:51:42&-73:13:47&NC&0.75&0.75& - &in H-A26 \& SMC-DEM60,MA485,OGLE72              \\
 29nw&B59,L61-183,MA488,        &0:51:44&-72:50:25&CN&0.80&0.60& 80&mP \& OGLE73\\
 29nw&H86-123,OGLE206       &0:51:44&-73:10:01& C&0.55&0.55& - &in BS51                                            \\
 29nw&H86-127,OGLE207       &0:51:49&-72:32:28& C&0.60&0.40& 20&in B-OB9                                           \\
 29nw&K29,L44,OGLE74        &0:51:53&-72:57:14& C&0.95&0.95& - &                                                   \\
 29nw&H86-126,OGLE75        &0:51:54&-73:05:53& C&0.50&0.50& - &                                                   \\
 29nw&H86-128,OGLE208       &0:52:03&-72:49:04& C&0.45&0.45& - &                                                   \\
 29nw&H86-129,OGLE76        &0:52:12&-72:31:51& C&0.65&0.45& 40&in SMC-DEM64                                       \\
 29nw&BS56,OGLE77           &0:52:13&-73:00:12& C&0.70&0.55& 90&mP                                                 \\
 29nw&BS253,OGLE209         &0:52:15&-72:45:58&AC&0.60&0.40& 10&                                                   \\
\tableline
\end{tabular}
\end{scriptsize}
\tablecomments{Table 1 continued.
}
\end{table*}

\clearpage

\begin{table}
\begin{scriptsize}
\renewcommand{\tabcolsep}{0.9mm}
\begin{tabular}{llccccccl}
\tableline\tableline &&&&&D$_{max}$&D$_{min}$&P.A.&\\
Plate&Name&R.A.(2000)&Dec(2000)&T&(arcmin)&(arcmin)&(deg)&Remarks\\
(1)&(2)&(3)&(4)&(5)&(6)&(7)&(8)&(9)\\ \tableline
29nw&H86-125,OGLE79        &0:52:17&-73:22:32&CA&0.75&0.50& 80&in SMC-DEM70n                                      \\
 29nw&H86-130,OGLE78        &0:52:17&-73:01:04& C&0.75&0.60&  0&mP                                                 \\
 29nw&B64,OGLE210           &0:52:30&-73:02:59& C&0.70&0.70& - &mP,in H-A29                                        \\
 29nw&BS57,OGLE211          &0:52:32&-73:02:10& C&0.65&0.45& 60&mP,in H-A29                                        \\
 29nw&H86-133,OGLE81        &0:52:34&-72:40:54& C&0.75&0.75& - &in H-A31                                           \\
 29nw&H86-132,OGLE80        &0:52:31&-72:37:46& C&0.50&0.40& 70&in SMC-N50                                         \\
 29nw&BS60,OGLE82           &0:52:42&-72:55:32& C&0.80&0.80& - &                                                   \\
 29nw&H86-134w,OGLE212      &0:52:45&-72:59:24& C&0.50&0.50& - &mT,in H-A30                                        \\
 29nw&B65,OGLE83            &0:52:44&-72:58:48& C&0.75&0.75& - &mT                                                 \\
 29nw&B66,OGLE85            &0:52:48&-72:47:46& C&0.60&0.45& 40&                                                   \\
 29nw&H86-134e,OGLE213      &0:52:48&-72:59:22& C&0.50&0.45&  0&mT,in H-A30                                        \\
 29nw&BS63,OGLE84           &0:52:47&-73:24:25& C&0.50&0.40&150&mP, in SMC-DEM73                                   \\
 29nw&B67,OGLE87            &0:52:49&-73:24:43& C&0.65&0.50&110&mP,in SMC-DEM73                                    \\
 29nw&H86-135,OGLE86        &0:52:48&-72:30:38& C&0.55&0.55& - &                                                   \\
 29nw&K31,L46,OGLE88        &0:53:01&-72:53:49& C&2.80&2.80& - &sup SMC-DEM69                                      \\
 29nw&B69,OGLE89            &0:53:07&-72:37:28& C&0.65&0.65& - &in H-A33                                           \\
 29nw&NGC294,L47,ESO29SC22,     &0:53:06&-73:22:49& C&1.70&1.70& - &\& OGLE90                                       \\
 29nw&H86-140,OGLE214       &0:53:09&-72:49:58& C&0.45&0.40& 50&mP                                                 \\
 29nw&H86-138,OGLE91        &0:53:10&-72:34:25& C&0.45&0.45& - &in H-A36                                           \\
 29nw&BS256,OGLE215         &0:53:17&-72:44:03& C&0.50&0.45& 80&                                                   \\
 29nw&B71,OGLE92            &0:53:18&-72:46:00& C&0.85&0.85& - &                                                   \\
 29nw&H86-143,OGLE93        &0:53:31&-72:40:04& C&0.80&0.80& - &m6,in H-A35                                        \\
 29nw&SMC-N52A,L61-243,         &0:53:40&-72:39:35&NC&0.50&0.50& - &m6,in HA35\& DEM77sw,MA696,OGLE94         \\
 29nw&SMC-N52B,L61-244,B73,     &0:53:42&-72:39:15&NC&0.50&0.50& - &m6,in HA35\& DEM77ne,MA699,OGLE96         \\
 29nw&BS68,OGLE95           &0:53:42&-73:21:32&CA&0.90&0.75&130&mP                                                 \\
 29nw&H86-147,OGLE216       &0:53:50&-72:53:47& C&1.20&1.20& - &in SMC-DEM69                                       \\
 29nw&BS69,OGLE217          &0:53:56&-72:51:24&CA&0.60&0.40& 45&mP                                                 \\
 29nw&BS72,OGLE97           &0:54:11&-72:51:54&CA&0.75&0.60& 20&mP                                                 \\
 29nw&H86-152,OGLE218       &0:54:23&-72:41:41& C&0.45&0.45& - &in B-OB11                                          \\
 29nw&B80,OGLE98            &0:54:47&-73:13:25& C&0.80&0.65&150&                                                   \\
 29nw&B79,OGLE99            &0:54:48&-72:27:58& C&0.80&0.60&100&in H-A38                                           \\
 29nw&B76,OGLE219           &0:54:52&-72:54:59& C&0.40&0.40& - &                                                   \\
 29nw&H86-159,OGLE102       &0:55:12&-72:41:00& C&0.50&0.40&130&mP,in BS260                                        \\
 29nw&H86-158,OGLE100       &0:55:09&-72:48:40& C&0.65&0.65& - &in B-OB12                                          \\
 29nw&H86-155,OGLE101       &0:55:12&-73:17:48& C&0.50&0.50& - &                                                   \\
 29nw&OGLE220               &0:55:14&-72:36:04&CA&0.35&0.35& - &                                                   \\
 29nw&B83,OGLE103           &0:55:30&-73:04:17& C&0.55&0.55& - &                                                   \\
 29nw&K34,L53,OGLE104       &0:55:33&-72:49:58& C&1.20&1.20& - &                                                   \\
 29nw&H86-165,OGLE105       &0:55:43&-72:52:48& A&1.10&1.10& - &                                                   \\
 29nw&H86-164,OGLE221       &0:55:45&-72:42:18& C&0.45&0.35& 70&in BS262                                           \\
 29nw&OGLE106               &0:56:09&-73:12:22&CA&0.85&0.85& - &                                                   \\
 29nw&NGC330,K35,L54,ESO29SC24, &0:56:19&-72:27:50& C&2.80&2.50&120&\& OGLE107 in H-A40,sup?SMC-DEM87               \\
 29nw&B86,OGLE222           &0:56:16&-72:30:59& C&0.65&0.65& - &                                                   \\
 29nw&BS81,OGLE223          &0:56:26&-72:29:45& C&0.60&0.55&  0&mP,in H-A40                                        \\
 29nw&H86-172,OGLE108       &0:56:34&-72:30:08& C&0.55&0.55& - &mP,in H-A40                                        \\
 29nw&OGLE224               &0:56:46&-72:45:06&CA&0.40&0.40& - &                                                   \\
 29nw&H86-174,OGLE225       &0:57:18&-72:56:01& C&0.45&0.45& - &                                                   \\
 29nw&L56,SMC-S26,OGLE109   &0:57:31&-72:15:52& C&0.95&0.95& - &                                                   \\
 29nw&H86-178,OGLE110       &0:57:46&-72:42:21& C&0.55&0.50& 70&                                                   \\
 29nw&BS88,OGLE111          &0:57:50&-72:56:37& C&0.55&0.50&135&                                                   \\
 29nw&H86-175,OGLE227       &0:57:50&-72:26:24& C&0.40&0.40& - &mP                                                 \\
 29nw&H86-177,OGLE226       &0:57:50&-72:30:29& C&0.75&0.75& - &mP,in B-OB13                                       \\
 29nw&H86-179,OGLE112       &0:57:57&-72:26:42& C&0.40&0.40& - &mP                                                 \\
 29nw&SMC-N63,L61-331,SMC-DEM94,&0:58:16&-72:38:47&NA&0.60&0.60& - &mT,in H-A42 \& MA1065,OGLE113                   \\
 29nw&H86-181,OGLE228       &0:58:19&-72:17:57& C&0.65&0.65& - &in H-A43                                           \\
 29nw&SMC-N64A,L61-335,SMC-DEM95,         &0:58:26&-72:39:57&NC&0.80&0.65& 70&mT,in N64 \& H86-182,MA1071,OGLE114\\
 29nw&H86-183,OGLE115       &0:58:34&-72:16:52& C&0.55&0.55& - &                                                   \\
 29nw&BS272,OGLE229         &0:58:38&-72:14:04&NC&0.65&0.65& - &mP,in SMC-DEM98                                    \\
 29nw&OGLE116               &0:59:05&-72:47:12&AC&1.00&0.70& 90&                                                   \\
 29nw&B96,OGLE117           &0:59:14&-72:36:29& C&1.00&0.90& 80&att SMC-DEM114                                     \\
 29nw&IC1611,K40,L61,ESO29SC27, &0:59:48&-72:20:02& C&1.50&1.50& - &m4 \& OGLE118                                   \\
 29nw&H86-186,OGLE119       &0:59:57&-72:22:24& C&0.60&0.60& - &m4,att SMC-DEM114                                  \\
\tableline
\end{tabular}
\end{scriptsize}
\tablecomments{Table 1 continued.
}
\end{table}

\clearpage

\begin{table}
\begin{scriptsize}
\renewcommand{\tabcolsep}{0.9mm}
\begin{tabular}{llccccccl}
\tableline\tableline &&&&&D$_{max}$&D$_{min}$&P.A.&\\
Plate&Name&R.A.(2000)&Dec(2000)&T&(arcmin)&(arcmin)&(deg)&Remarks\\
(1)&(2)&(3)&(4)&(5)&(6)&(7)&(8)&(9)\\ \tableline
29nw&IC1612,K41,L62,ESO29SC28, &1:00:01&-72:22:08& C&1.20&0.80& 20&m4,att SMC-DEM114 \& OGLE120                    \\
 29nw&H86-188,OGLE121       &1:00:13&-72:27:44&AC&1.30&0.70& 50&in SMC-DEM114                                      \\
 29nw&B99,OGLE122           &1:00:27&-73:05:12& C&0.75&0.75& - &                                                   \\
 29nw&H86-189,OGLE123       &1:00:33&-72:14:23& C&0.40&0.40& - &mP                                                 \\
 29nw&H86-190,OGLE230       &1:00:33&-72:15:31& C&0.40&0.40& - &mP                                                 \\
 29nw&K42,L63,OGLE124       &1:00:34&-72:21:56& C&0.85&0.85& - &m4,att SMC-DEM114                                  \\
 29nw&OGLE125               &1:00:47&-72:55:41&CA&0.80&0.70& 10&                                                   \\
 29nw&H86-191,OGLE231       &1:00:58&-72:32:25& C&0.80&0.80& - &mP,in? SMC-DEM114                                  \\
 29nw&L65,H86-192,OGLE126   &1:01:02&-72:45:05& C&1.10&1.10& - &                                                   \\
 29nw&H86-194,OGLE232       &1:01:14&-72:33:03& C&0.85&0.85& - &mP,in? SMC-DEM114                                  \\
 29nw&H86-193,OGLE127       &1:01:18&-72:13:42& C&0.55&0.55& - &in B-OB19                                          \\
 29nw&B105,OGLE128          &1:01:37&-72:24:25& C&0.75&0.75& - &in?SMC-DEM114                                      \\
 29nw&L66,OGLE129           &1:01:45&-72:33:52& C&1.10&1.10& - &att SMC-DEM114                                     \\
 29nw&B108,OGLE130          &1:01:52&-72:10:58& C&0.80&0.80& - &in B-OB19                                          \\
 29nw&OGLE131               &1:02:03&-72:31:19&CA&0.70&0.70& - &                                                   \\
 29nw&OGLE132               &1:02:13&-72:57:59& C&0.65&0.65& - &                                                   \\
 29nw&OGLE133               &1:02:31&-72:19:06& C&0.80&0.80& - &                                                   \\
 29nw&BS106,OGLE233         &1:02:40&-72:23:50&NC&0.55&0.55& - &in SMC-DEM118, mP                                  \\
 29nw&B114,OGLE234          &1:02:53&-72:24:53&NC&0.85&0.85& - &in SMC-DEM118,mP                                   \\
 29nw&K47,L70,OGLE134       &1:03:12&-72:16:21& C&0.75&0.75& - &                                                   \\
 29nw&OGLE135               &1:03:17&-72:44:27&AC&1.45&1.45& - &                                                   \\
 29nw&OGLE136               &1:03:22&-72:27:57&CA&0.80&0.50&100&                                                   \\
 29nw&B115,OGLE137          &1:03:23&-72:39:06& C&0.90&0.80&100&in H-A52                                           \\
 51se&OGLE138               &1:03:53&-72:06:11&CA&0.60&0.60& - &mP,in H-A53                                        \\
 29nw&NGC376,K49,L72,ESO29SC29, &1:03:53&-72:49:34& C&1.80&1.80& - &mP \& OGLE139                                   \\
 29nw&BS114,OGLE235         &1:03:59&-72:48:18&AC&0.70&0.50&110&mP                                                 \\
 29nw&OGLE144,OGLE236   &1:04:05&-72:07:15&CA&0.60&0.60& - &mP                                                 \\
 29nw&BS118,OGLE140         &1:04:14&-72:38:49& A&1.30&0.90&120&                                                   \\
 29nw&BS121,OGLE237         &1:04:22&-72:50:52& C&1.60&1.30&140&                                                   \\
 29nw&B121,OGLE141          &1:04:30&-72:37:09& C&0.65&0.65& - &                                                   \\
 51se&K50,L74,ESO51SC15,        &1:04:36&-72:09:38& C&1.00&1.00& - &in H-A54 \& OGLE142                             \\
 29nw&BS125,OGLE143         &1:04:40&-72:33:00& A&1.60&0.85& 30&                                                   \\
 51se&SMC-N78B,L61-439,         &1:05:04&-71:59:25&NC&0.40&0.30&100&m5,in NGC395 \& MA1508/1514,OGLE145             \\
 51se&MA1520,OGLE147        &1:05:08&-71:59:45&NC&0.50&0.45&130&m5,in NGC395                                       \\
 29nw&OGLE148               &1:05:10&-72:36:08&CA&0.95&0.95& - &                                                   \\
 51se&OGLE146               &1:05:13&-71 59 42&NA&0.45&0.45&   &m5,in NGC395                                       \\
 51se&IC1624,K52,L76,ESO51SC17, &1:05:22&-72:02:35& C&0.90&0.90& - &in SMC-N78 \& OGLE149                           \\
 29nw&BS131,OGLE150         &1:06:00&-72:20:29& C&0.55&0.50&150&in B-OB29                                          \\
 29nw&OGLE151               &1:06:13&-72:47:39&AC&1.20&1.20& - &                                                   \\
 29nw&OGLE152               &1:06:21&-72:49:48&CA&0.65&0.65& - &                                                   \\
 29nw&K54,L79,ESO29SC31,        &1:06:48&-72:16:25& C&0.90&0.90& - &in B-OB29  \&OGLE153                            \\
 29nw&B129,OGLE154          &1:07:02&-72:37:18&CA&1.10&1.10& - &                                                   \\
 29ne&K56,OGLE155           &1:07:28&-72:29:36& C&1.10&0.90&130&                                                   \\
 29ne&L80,OGLE156           &1:07:28&-72:46:10& C&1.20&1.20& - &                                                   \\
 29ne&K55,L81,OGLE157       &1:07:32&-73:07:11& C&0.95&0.95& - &                                                   \\
 29ne&OGLE238               &1:07:52&-72:31:55&CA&0.35&0.35& - &                                                   \\
 29ne&NGC416,K59,L83,ESO29SC32, &1:07:59&-72:21:20& C&1.70&1.70& - &\& OGLE158                                      \\
 29ne&NGC419,K58,L85,ESO29SC33, &1:08:19&-72:53:03& C&2.80&2.80& - &\& OGLE159                                      \\
 29ne&OGLE160               &1:08:37&-72:26:21&CA&0.65&0.65& - &                                                   \\
 29ne&K61,OGLE161           &1:09:03&-73:05:12& C&0.95&0.95& - &                                                   \\
\tableline
\end{tabular}
\end{scriptsize}
\tablecomments{Table 1 continued.
}
\end{table}

\clearpage

\begin{table}
\begin{center}
\caption[]{Distribution of OGLE objects in different catalogs}
\begin{scriptsize}
\renewcommand{\tabcolsep}{0.9mm}
\begin{tabular}{lccc}
\tableline\tableline Acronym&Reference&Entries&Entries\\
&&Pietrzy\'nski et al. 1998&This paper\\ \hline NGC&Dreyer
1888&16&18\\ IC&Dreyer 1895&3&3\\ K&Kron 1956&12&12\\ SMC-N&Henize
1956&--&7\\ L&Lindsay 1958&8&9\\ L61&Lindsay 1961&7&--\\ HW&Hodge
\& Wright 1974&3&4\\ B&Br\"uck 1976&33&37\\ H86&Hodge 1986&60&69\\
BS&Bica \& Schmitt 1995&24&32\\ MA&Meyssonnier \& Azzopardi
1993&--&1\\ SMC-OGLE&Pietrzy\'nski et al. 1998&72&46\\ \tableline
\end{tabular}
\end{scriptsize}
\end{center}
\end{table}

\clearpage

\begin{table}
\caption[]{Updated SMC/Bridge catalog including SMC\_OGLE
objects}
\begin{scriptsize}
\renewcommand{\tabcolsep}{0.9mm}
\begin{tabular}{llccccccl}
\tableline\tableline &&&&&D$_{max}$&D$_{min}$&P.A.&\\
Plate&Name&R.A.(2000)&Dec(2000)&T&(arcmin)&(arcmin)&(deg)&Remarks\\
(1)&(2)&(3)&(4)&(5)&(6)&(7)&(8)&(9)\\ \tableline 28nw&
AM-3,ESO28SC4& 23:48:59&  -72:56:43&      C& 0.90& 0.90&  -&\\
28ne& L1,ESO28SC8& 0:03:54&  -73:28:19&      C& 4.60& 4.60&  -&\\
28ne& L2& 0:12:55& -73:29:15&      C& 1.20& 1.20&  -&\\ 28ne& HW1&
0:18:25& -73:23:38&     CA& 0.95& 0.85&   0&\\ 28ne& L3,ESO28SC13&
0:18:25& -74:19:07&      C& 1.00& 1.00&  -&\\ 28ne& BS1& 0:18:55&
-73:24:52&     CA& 0.90& 0.90&  -&\\ \tableline
\end{tabular}
\end{scriptsize}
\tablecomments{This table is published in its entirety in the
electronic edition of The Astronomical Journal.}
\end{table}

\clearpage

\begin{table}
\caption[]{Census of object types in the Updated SMC catalog}
\begin{scriptsize}
\renewcommand{\tabcolsep}{0.9mm}
\begin{tabular}{lcrclcc}
\tableline\tableline Types&(whole SMC)&&(OGLE Survey
area)&&(Outside&OGLE area)\\ (1)&(2)&(3)&(4)&(5)&(6)&(7)\\
\tableline C&460&~~~~251&162&14~~~~&209&12\\
CA&115&~~~~64&40&24~~~~&51&19\\ CN&9&~~~~6&5&--~~~~&3&--\\
A&154&~~~~75&4&1~~~~&79&1\\ AC&58&~~~~22&10&5~~~~&36&8\\
AN&40&~~~~26&--&--~~~~&14&--\\ NC&77&~~~~51&11&1~~~~&26&1\\
NA&198&~~~~126&6&1~~~~&72&1\\ N&11&~~~~10&--&--~~~~&1&--\\
Total&1122&~~~~631&238&46~~~~&491&42\\ \tableline
\end{tabular}
\end{scriptsize}
\tablecomments{ By columns: (1) Object type; (2) All objects up to
$\alpha = 2^h$;
 (3)  All objects in the OGLE area;
 (4)  All objects in the OGLE catalog (according to Table 1);
 (5)  All intrinsically new OGLE objects;
 (6)  All objects outside the OGLE region;
 (7)  Predicted CCD objects outside the OGLE region (see assumptions in the text).
}
\end{table}

\end{document}